\documentclass[aip,apl,preprint,showpacs]{revtex4-1}
\usepackage{graphicx}
\usepackage{graphics}

\begin{document}
	
	\title{Magnetic field tunable superconducting transition in Nb/Co/Py/Nb exchange spring multilayers}
	
	\author{Ekta Bhatia}
	\email[Ekta Bhatia: ]{bhatiaekta@niser.ac.in}
	\affiliation{School of Physical Sciences, National Institute of Science Education and Research (NISER), HBNI, Bhubaneswar, Odisha, India-752050}
	\author{J. M. Devine-Stoneman}
	\affiliation{Department of Materials Science \& Metallurgy, University of Cambridge, 27 Charles Babbage Road, CB3 0FS, United Kingdom}
	\author{Zoe H. Barber}
	\affiliation{Department of Materials Science \& Metallurgy, University of Cambridge, 27 Charles Babbage Road, CB3 0FS, United Kingdom}
	
	\author{J. W. A. Robinson }
	\affiliation{Department of Materials Science \& Metallurgy, University of Cambridge, 27 Charles Babbage Road, CB3 0FS, United Kingdom}

	\author{Kartik Senapati}
	\affiliation{School of Physical Sciences, National Institute of Science Education and Research (NISER), HBNI, Bhubaneswar, Odisha, India-752050}
	
	\date{\today}
	
	\begin{abstract}
Over the last decade it has been shown that magnetic non-collinearity at a \textit{s}-wave superconductor/ferromagnet interface is a key ingredient for spin-singlet to spin-triplet pair conversion. This has been verified in several synthetic non-collinear magnetic structures. A magnetically soft and hard ferromagnetic layer combination in a bilayer structure can function as a field tunable non-collinear magnetic structure which may offer magnetic-field tuneability of singlet-to-triplet pair conversion. From magnetization measurements of Nb/Co/Py/Nb multilayers we demonstrate a reversible enhancement of the superconducting critical temperature of 400 mK by measuring $T_{\mathrm c}$ with and without a non-collinear magnetic structure between Co and Py. The sensitivity of $T_{\mathrm c}$ in these structures offers the potential for realizing magnetic field tunable Josephson junctions in which pair conversion and Josephson critical currents are controllable using modest magnetic fields.
		
	\end{abstract}

	\maketitle

The field of superconducting spintronics has attracted significant interest in recent years\cite{1,31,9} with the aim of
creating fast, energy efficient logic and memory devices that operate in the superconducting state. By combining superconductivity (S) and ferromagnetism (F), interesting phenomena can arise, including $\pi$-Josephson coupling\cite{2} and spin-triplet supercurrents\cite{3}. For a homogeneous ferromagnet, spin-mixing  occurs at the S/F interface, giving rise to $S_{\mathrm z}$ = 0 triplet components\cite{9}. By introducing magnetic non-collinearity at the S/F interface, spin rotation occurs which converts the $S_{\mathrm z}$ = 0 triplet component into $S_{\mathrm z}$ = $\pm$1 triplet components\cite{9}. The $S_{\mathrm z}$ = $\pm$1 components are odd in frequency and even in momentum and, therefore, are insensitive to impurity scattering\cite{9}. When propagating through a ferromagnet, the Zeeman field has no pair-breaking effect on triplet Cooper-pairs meaning triplet Cooper pairs are long-ranged in the ferromagnet\cite{9}.

 A number of proposals have been put forward to create and control triplet supercurrents in S/F hybrids, including Josephson junctions with domain walls or textured ferromagnets \cite{10,11}, bilayer and trilayer ferromagnetic regions \cite{12}, spin injection \cite{13}, and via spin-active interfaces \cite{14}, where a net interface magnetic moment is misaligned with respect to the bulk magnetization. 
 
 The first demonstration of long-ranged supercurrents was reported by Keizer \textit{et al.}\cite{15} in 2006 through the observation of supercurrents through the half-metallic ferromagnet CrO$_{2}$. The results were later repeated by Anwar \textit{et al.}\cite{16}. In 2010, a series of experiments by different groups demonstrated spin-triplet pairing in Josephson junctions. Khaire \textit{et al.}\cite{17} used ferromagnetic/non-magnetic multilayer spin-mixers while Robinson \textit{et al.} \cite{18} used the helical rare-earth antiferromagnet Ho to generate triplet supercurrents in Co whilst Sprungmann \textit{et al.}\cite{19} used a Heusler alloy to generate the triplet supercurrent. All of these experiments share similarities to the SF$'$FF$'$S device proposed by Houzet and Buzdin \cite{20} where the F$'$/F interfaces are magnetically non-parallel.

Although it is now established that triplet supercurrents exist, practical application in superconducting spintronics requires direct control over the generation and tuning of triplet supercurrents. Hard/Soft bilayer interfaces with magnetic non-collinear structure have the potential to offer such control\cite{33}. Such interfaces consist of neighboring layers of magnetically hard and soft ferromagnetic materials in which the magnetic configuration is programmable using relatively small ($<$100 Oe) magnetic fields\cite{25}. A recent report \cite{21} on proximity effects at Nb/NiFe/SmFe (where NiFe/SmFe is a soft/hard bilayer interface with magnetic non-collinear structure) shows an enhancement in the superconducting critical temperature ($T_{\mathrm c}$) of $\sim$30 mK due to triplet pair formation. Zhu \textit{et al.}\cite{22} reported enhancements in $T_{\mathrm c}$ of $\sim$10 mK in Nb/NiFe/SmCo (where NiFe/SmCo is a soft/hard bilayer interface with magnetic non-collinear structure) multilayers through resistance measurements. Some other reports demonstrate\cite{26,27,34,35} a suppression in $T_{\mathrm c}$ in S/F$_{1}$/F$_{2}$ systems due to generation of triplet Cooper pairs in transport measurements.

In this paper, we investigate changes in $T_{\mathrm c}$ through magnetization measurements of Nb/Co/Py/Nb multilayers by varying the magnetic configuration of the Co/Py hard/soft bilayer interface. Diamagnetic currents are set up in Nb due to an in-plane applied magnetic field. The conversion of singlet pairs to triplet pairs and vice-versa at the interface of superconductor/non-collinear magnetic structure may therefore be probed via a modification in the diamagnetic current distribution in Nb. 

	For a Nb/Co/Py/Nb multilayer, we observe a decrease in $T_{\mathrm c}$ in the range of fields where hysteresis loop opens up. We explain this decrease due to the stray fields from intrinsic domain walls\cite{30}. However, in the range of fields where a relative angle between the Co and Py layer moments is established, we observe a gradual recovery in $T_{\mathrm c}$. We explain this recovery in $T_{\mathrm c}$ as being related to singlet-to-triplet pair conversion. Furthermore, the tunability of $T_{\mathrm c}$ is achieved in the magnetically reversible non-collinear regime of Co/Py.

Nb/Co/Py/Nb multilayers were prepared at room temperature by DC magnetron sputtering in an Ar pressure of 1.5 Pa onto square 5 $\times$ 5 mm$^{2}$ oxidized single crystal silicon substrates. The base pressure of the deposition chamber was of the order of $10^{-9}$ mBar. The thickness of Py was varied from 30 nm to 90 nm in steps of 15 nm while the thickness of Nb and Co layers were fixed at 55 nm and 30 nm, respectively. Magnetization measurements were performed in a Quantum Design SQUID-VSM with magnetic fields applied parallel to the plane along the film edge. The superconducting critical temperature ($T_{\mathrm c}$) was determined via measurements of magnetization vs temperature at different field values in the field range from positive to negative saturation. \textit{M}(\textit{T}) Measurements were performed using the ultra-low field (ULF) option of SQUID magnetometer. The ULF option uses a custom fluxgate field sensor in conjunction with in-situ concentric modulation and trim coils to actively cancel the residual magnetic field in the superconducting solenoid. The fluxgate sensor measures the residual field along the solenoid’s longitudinal axis and then nulls it by setting a DC field using compensation coils installed in the superconducting solenoid. After the flux cancellation, the sample is loaded in the SQUID magnetometer. All the measurements for a particular sample are performed in a matching loading geometry at same angle of inclination. Therefore, all the extraneous parameters that can affect the $T_{\mathrm c}$ remains fixed.

 Figure 1 shows a schematic illustration of the sample geometry along with \textit{M}(\textit{H}) loops of a Nb/Co(30 nm)/Py(30 nm)/Nb multilayer. The Nb layers are both 55 nm thick. Co and Py were chosen since they are magnetically hard and soft ferromagnetic materials so as to achieve a magnetically non-collinear structure with relatively small in-plane magnetic fields\cite{28,29}. The Co/Py interface is exchange coupled; however, Py is free to rotate with an external magnetic field at Nb/Py interface. Therefore, when a magnetic field is applied in a direction opposing the net magnetization [Fig. 1(a)] a relative angle between the Co and Py moments forms. This magnetization angle results in a non-collinear magnetic structure that is controllable with in-plane magnetic field of a few Oe. Furthermore, this non-collinear structure is magnetically reversible in a particular field range, termed as reversible range, similar to an exchange-spring\cite{25}.
 
 \begin{figure}[htbp!]
 	\includegraphics[width=7cm]{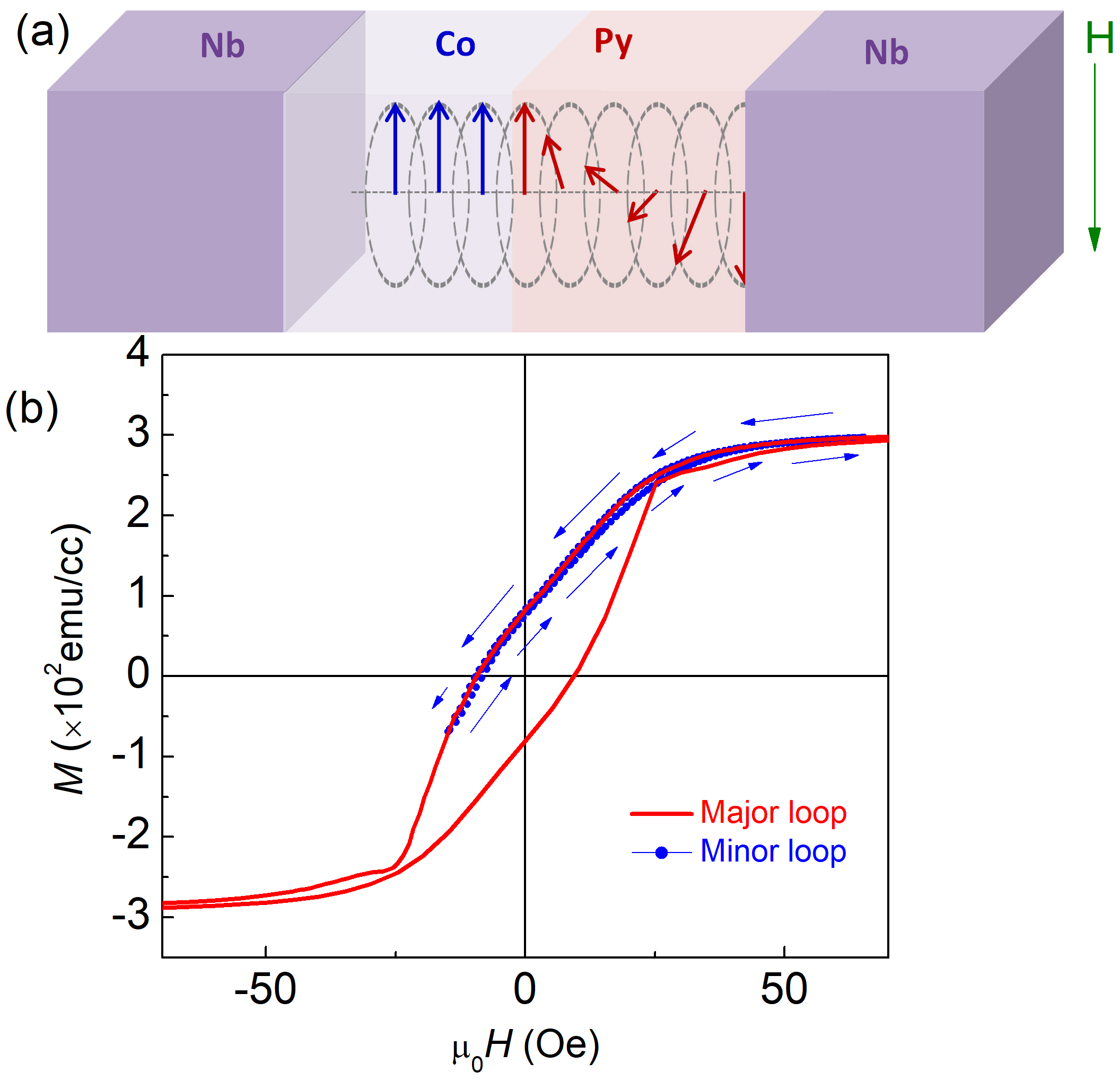}\\
 	\caption{(a) Schematic illustration showing magnetization rotation in Nb/Co/Py/Nb multilayer in which a non-collinear magnetic structure forms in Co/Py bilayer. (b) Major and minor \textit{M}(\textit{H}) loops of a Nb/Co(30 nm)/Py(30 nm)/Nb multilayer at 10 K in an in-plane magnetic field.}
 \end{figure}
  
 Figure 1(b) shows the major and minor \textit{M}(\textit{H}) loops of Nb/Co(30 nm)/Py(30 nm)/Nb multilayer at 10 K. In the major loop, magnetic field is varied from positive to negative saturation and then reversed back to positive saturation. In the minor loop, the magnetic field is varied from positive saturation to a field that is less than negative saturation and then reversed back to positive saturation. The minor loop is found to be reversible upto -14 Oe, as shown in Fig. 1(b). This shows that the non-collinearity formed in Py can be tuned with an applied magnetic field of a few tens of Oe in Co(30 nm)/Py(30 nm) bilayer\cite{23,24,25}. 
  	
\begin{figure}[htbp!]
	\centering
	\includegraphics[width=8cm]{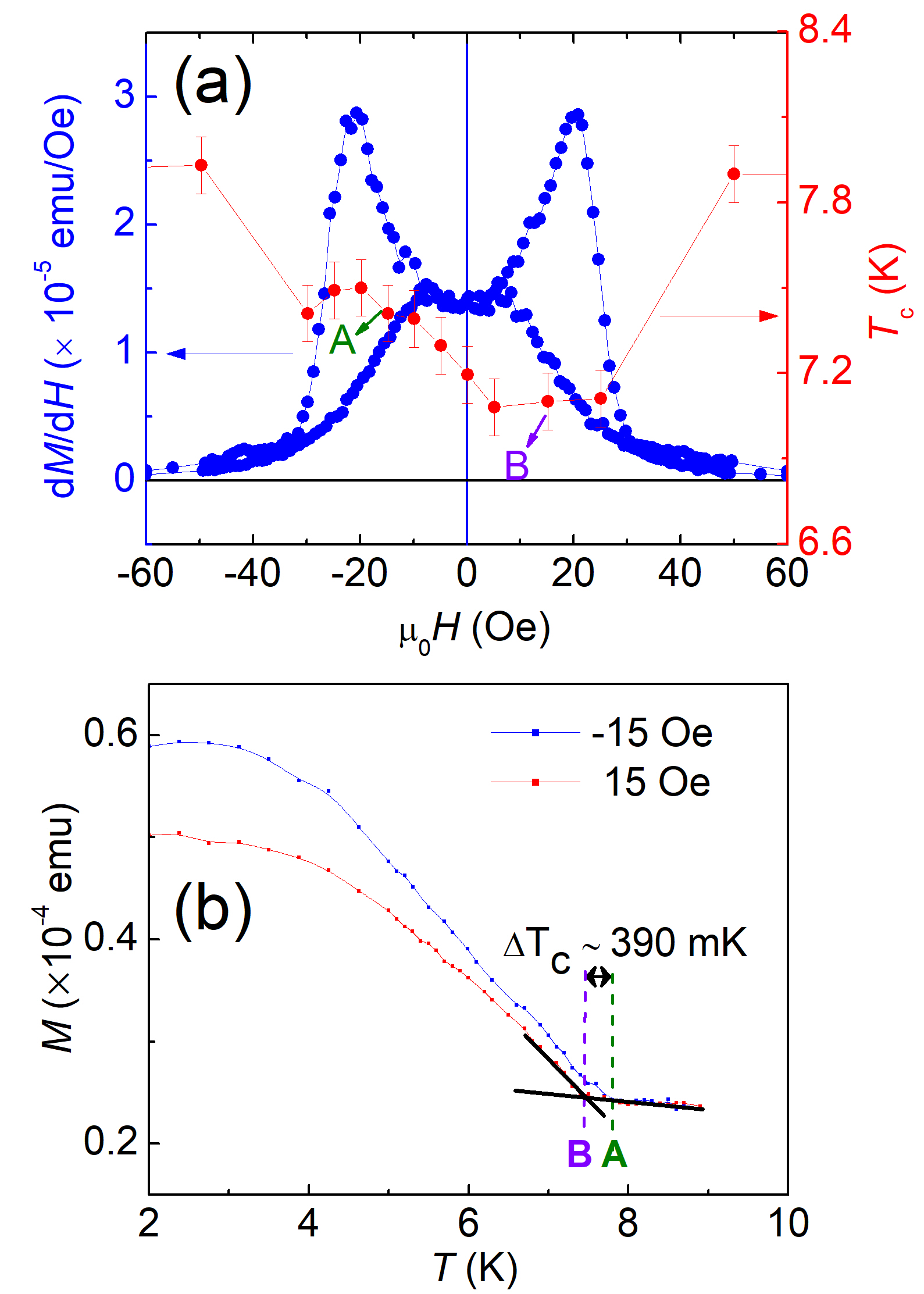}\\

	\caption{(a) $T_{\mathrm c}(H)$ (red points; extracted from \textit{M}(\textit{T}) measurements, right hand axis) and d\textit{M}/d\textit{H} curve (blue points, left hand axis) for a Nb/Co(30 nm)/Py(30 nm)/Nb multilayer. (b) Representitive \textit{M}(\textit{T}) curves [A and B in (a)] from which $T_{\mathrm c}$ is estimated. The curves are shifted along the y-axis to enable a clearer comparison of the two curves.  }
\end{figure}	
    In Figure 2(a) we have plotted $T_{\mathrm c}(H)$ curves for a Nb/Co(30 nm)/Py(30 nm)/Nb multilayer along with d$M$/d$H$ vs applied field. The d$M$/d$H$ curve is plotted by taking the derivative of magnetization (M) from \textit{M}(\textit{H}). To emphasize the change in $T_{c}$, we have plotted \textit{M}(\textit{T}) curves at two different magnetic fields: A and B of Fig. 2(a), as shown in Fig. 2(b). The $T_{\mathrm c}$ is defined as the point in \textit{M}(\textit{T}) where the superconducting state initiates and M starts to increase above the baseline shown by black solid line in Fig. 2(b). The $T_{\mathrm c}$ for the two curves A and B are marked by the dotted lines shown in green and purple color in Fig. 2(b). Each \textit{M}(\textit{T}) measurement in Fig. 2(b) has been taken by cooling the sample in zero magnetic field. Prior to each measurement, the sample is saturated in 200 Oe at 2 K and then the magnetic field is ramped to the measurement field at the same temperature. Notice that the magnetic moment in the superconducting state in this figure is positive, unlike the usual negative moment due to a diamagnetic response of Nb. Usually, in the superconducting state of a Nb film diamagnetic currents are set up in response to an in-plane applied magnetic field. This makes the magnetization of the superconducting state to drop below the magnetization value of the normal state at the transition temperature. However, in some of our \textit{M}(\textit{T}) measurements of Nb/Co/Py/Nb multilayers, we observe an increase in magnetization in the superconducting state compared to the value in the normal state, as shown in Fig. 2(b) and Fig. 3(a). Such increase in magnetization in the superconducting state can arise from the magnetic field history involved in the measurement, following the M(H) curve of superconducting Nb. Prior to each measurement, the sample was saturated in 200 Oe at 2 K and then the magnetic field was ramped to the measurement field at the same temperature. For temperatures below superconducting transition, the nature of the \textit{M}(\textit{T}) curve may be different due to field history despite the fact that the superconducting transition temperature remains unchanged. In the $T_{\mathrm c}(H)$ curve in Fig. 2(a), when the magnetic field is ramped down from saturation, domain formation in Py starts around the field where d$M$/d$H$ starts to increase, as shown in Fig. 2(a). A decrease in transition temperature with decreasing magnetic field is observed in this magnetic field range. This decrease in $T_{\mathrm c}$ can be explained due to the effect of stray fields from intrinsic domain walls in Py. The stray fields are present throughout the field range from positive to negative magnetic saturation and act to decrease $T_{\mathrm c}$ during the magnetization reversal of Co and Py\cite{30}. However, we observe a gradual recovery of $T_{\mathrm c}$ of about 400 mK in the range of fields where a relative angle between the Co and Py layer moments is established, as shown in Fig. 2(a). In this range of magnetic fields, the singlet pairs may convert into triplet pairs due to the non-collinear structure of magnetic moments between Co and Py. As a result, a modification in $T_{\mathrm c}$ happens in the non-collinear range of magnetic field. For temperatures above superconducting transition, the comparison of magnetizations in M(T) are in agreement with M(H). For temperatures below superconducting transition, the nature of the M(T) curve may be different due to field history despite the fact that the superconducting transition temperature remains unchanged.

    Figure 2(a) shows that dM/dH starts to rise below 40 Oe as the applied magnetic field is reduced from saturation, indicating the nucleation of domain walls in Py. Once the Py domains start to rotate around a field of 30 Oe, the non-collinear magnetic structure forms in Py with a corresponding change in magnetic moment – i.e. d$M$/d$H$ sharply rises. The non-collinear range is therefore defined from the field where d$M$/d$H$ sharply rises in Fig. 2(a) (matching with the opening of hysteresis loop and also opening of d$M$/d$H$ loop), indicating the rotation of Py domains. This range spans from +30 Oe to -30 Oe. In this field range we see a consistent recovery of $T_{\mathrm c}$ from 7.1 K. The initial drop in $T_{\mathrm c}$ from 7.9 K to 7.1 K is thus outside the non-collinear range of magnetic fields.   
    
    We emphasize that in a thin film geometry, the Co and Py layers are multidomain with a distribution of domain sizes. From magnetization vs in-plane magnetic field \textit{M}(\textit{H}) loops (and from the calculated d$M$/d$H$ curves) it is clear that neither the Co nor the Py magnetizations sharply switch direction as the magnetic field sweeps from positive to negative saturation. This is essentially due to the multidomain nature of Py and Co. Therefore, we should expect a degree of magnetic non-collinearity between Co and Py during the magnetization reversal of Py and  Co. During rotation of Py domains (while Co layer has not started rotating), the Nb/Py interface is magnetically inhomogeneous which favours singlet-to-triplet pair conversion due to diffusion of diamagnetic current across the interface. Similarly, during the gradual rotation of Co domains (while Py domains have already reversed) the Nb/Co interface is effective for singlet-triplet conversion.    
    In our experiment we are probing the superconductor proximity effect and singlet-to-triplet pair conversion via a modification in the diamagnetic current distribution in the Nb layers. Previous experiments\cite{21,22} investigate pair conversion via critical temperature ($T_{\mathrm c}$) measurements, where shifts in $T_{\mathrm c}$ are determined through changes in electronic resistance with magnetic field. Such measurements, however, are not volumetric since only the highest $T_{\mathrm c}$ within the S layer is extracted - i.e. currents shunt to regions in Nb with the highest $T_{\mathrm c}$.

  A recent report\cite{21} on Nb/Py/SmFe multilayers shows an enhancement of $T_{\mathrm c}$ of  about 30 mK in transport measurements. This is explained as due to the odd-triplet superconductivity generation through the non-collinear structure of Py/SmFe bilayer. Another report\cite{22} on Nb/Py/SmCo multilayers demonstrates a decrease in resistance ($\bigtriangleup$R) of $\sim$35\% in the non-collinear magnetic range of Py/SmCo bilayer. Here, the Py/SmCo bilayer thin films are single crystal and the authors claim a well-defined non-collinear structure compared to the previous report\cite{21} where Py/SmFe bilayer thin films were grown in polycrystalline form. Their calculations show that the $\bigtriangleup$R is equivalent to an enhancement in $T_{\mathrm c}$ of $\sim$10 mK in transport measurements. They propose that  this enhancement is due to an unanticipated proximity effect and there may be some interesting physics yet to be captured. In our study, magnetization measurements have been performed to extract the value of $T_{\mathrm c}$, and the modification of diamagnetic current distribution of Nb, due to singlet-triplet conversion, has been proposed as the mechanism for the recovery of $T_{\mathrm c}$ in the inhomogeneous range of Co and Py moments. Moreover, as the magnetic field approaches negative saturation, $T_{\mathrm c}$ again recovers to the value at positive saturation as shown in Fig. 2(a). 
  
  The coercive fields of Co and Py are indeed similar. The key ingredient to induce singlet-to-triplet pair conversion is a magnetically non-collinear structure at a superconductor(S)/ferromagnet(F) interface. With Co/Py, we obtain a non-collinear magnetic structure as inferred from the dM/dH vs H curve in Fig. 2(a). According to Baker et al. \cite{33}, Co/Py is a good exchange-spring candidate for generating triplet pairs in magnetic Josephson junctions. An additional benefit to Co/Py over SmCo/Py relates to the low coercive field of the exchange-spring interface, which enables triplet pair creation with relatively small (mT) magnetic fields, which do not affect the properties of the superconducting layers. For SmCo/Py bilayers, the large coercive fields and hardness of SmCo layer can pose a challenge to achieve triplet Josephson currents and large delta $T_{\mathrm c}$\cite{33,Hedge}.

Figure 3(a) shows the \textit{M}(\textit{T}) curves for different values of returning field, $H_{\mathrm ret}$, but the same measurement field, $H_{\mathrm meas}$, of 11 Oe. The sample was first saturated to 200 Oe at 2 K and then ramped to the returning field value, $H_{\mathrm ret}$, and then ramped to 11 Oe for each \textit{M}(\textit{T}) measurement. It was found that, when the returning field values lie in the reversible range, $T_{\mathrm c}$ remains the same, as shown in Fig. 3(a). But, if the returning field values are outside the reversible range, $T_{\mathrm c}$ is changed, as shown by the dotted lines in Fig. 3(a). 
\begin{figure}[htbp!]
	\includegraphics[width=8cm]{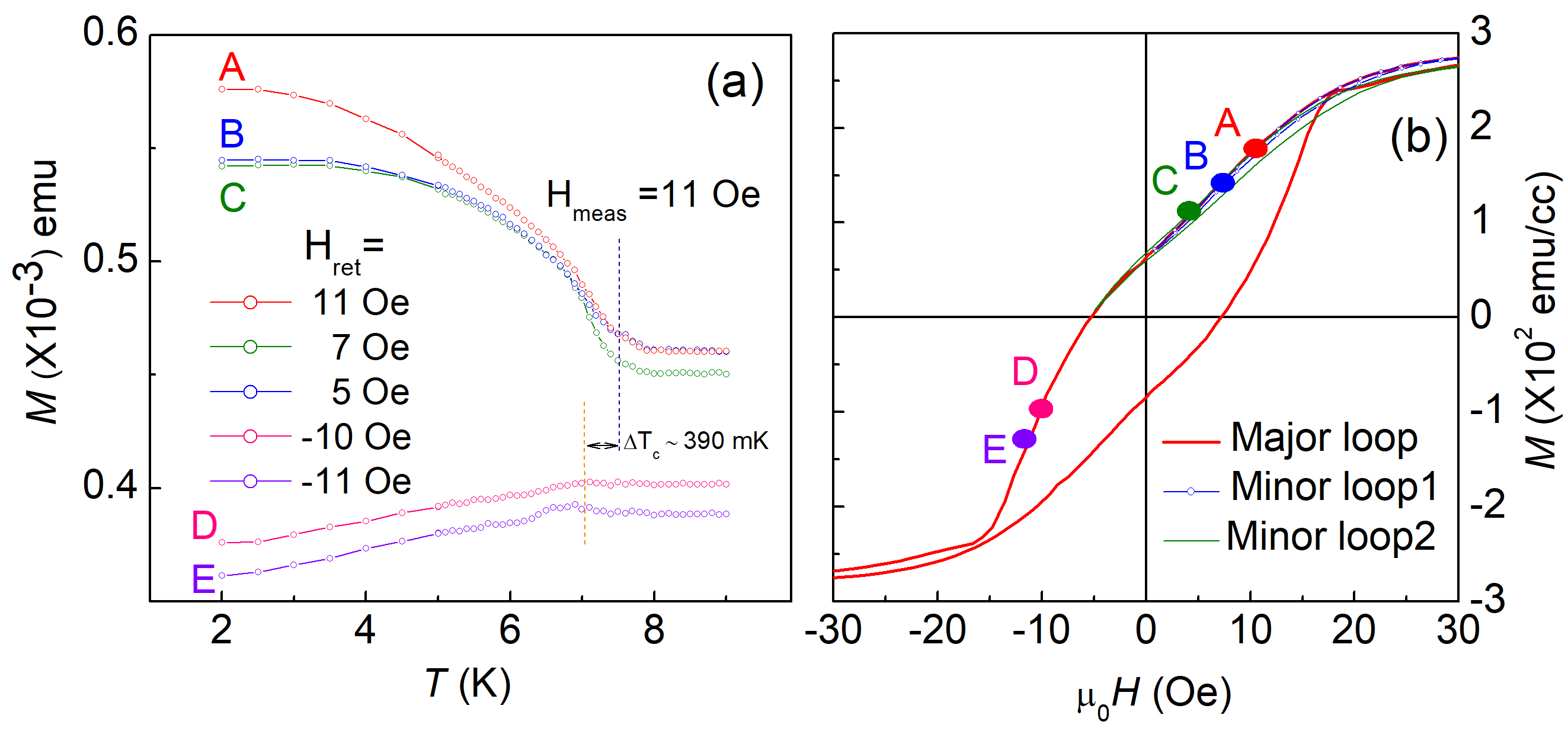}\\
	\caption{\textit{M}(\textit{T}) curves measured at 11 Oe with different returning fields H$_{\mathrm ret}$, illustrating the tuning of $T_{\mathrm c}$ in the reversible range of magnetic field. The returning fields are marked A, B, C, D, and E on the \textit{M}(\textit{H}) loops in panel (b).}
\end{figure}

The $T_{\mathrm c}$ above the loop region was 7.7 K for the Co(30 nm)/Py(90 nm) sample. On reversing from a field outside the reversible range (points D and E) in Fig. 3(a), the minor loops will be hysteretic with a net magnetic moment (at the measurement field) that is smaller than the non-hysteretic minor loops. This is because some of the Co domains that have rotated in the negative field direction (for points D and E) cannot switch back to the positive field (measurement field at point A) direction. Therefore, there is a net reduction in magnetic non-collinearity whilst sweeping the field from points D and E to A in Fig. 3(b). We observe in Fig. 3(a) that the reduced magnetic non-collinearity results in a reduction of $T_{\mathrm c}$ from about 7 K to 7.5 K for non-hysteretic returning fields B and C.   
   
\begin{figure}[htbp!]
	\includegraphics[width=7cm]{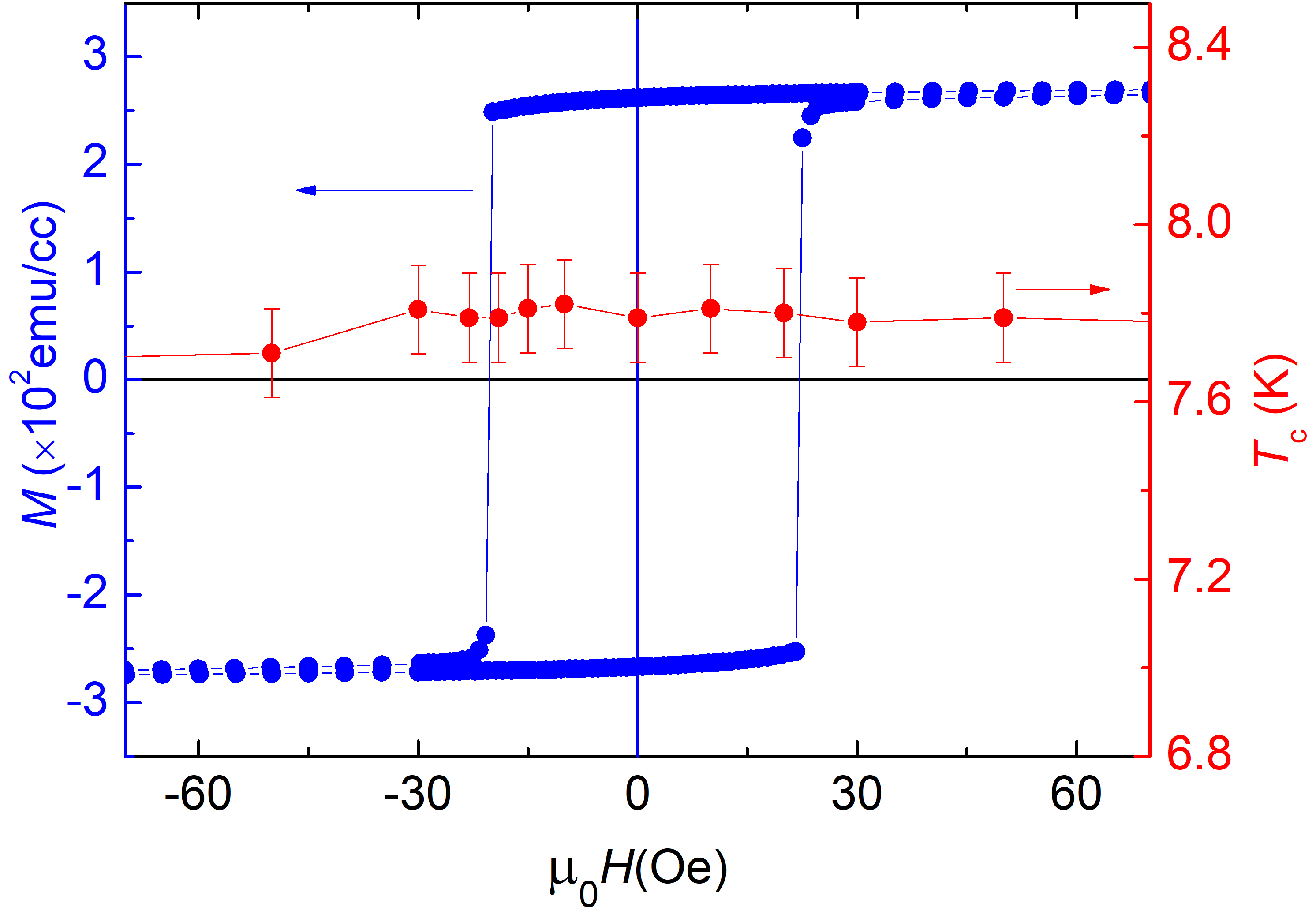}\\
	\caption{$T_{\mathrm c}(H)$ curve (red points, on the right hand axis) for Nb/Co(30)/Py(45)/Nb along with the corresponding \textit{M}(\textit{H}) loop (blue points, on the left hand axis). The square \textit{M}(\textit{H}) loop for Co(30)/Py(45) bilayer indicates a collinear magnetic configuration between Co and Py with no changes in the $T_{\mathrm c}$.}
\end{figure}

To investigate whether these results are due to singlet-triplet pair conversion, we measured a Nb/Co(30 nm)/Py(45 nm)/Nb multilayer for which non-collinear magnetic structure is not formed, as shown in Fig. 4. Earlier reports\cite{23,24,25} have shown that the hard layer/soft layer combination can act as an exchange spring with a non-collinear magnetic structure for certain thickness combinations, whilst for others, it can act like as a single magnetic layer. Fig. 4 shows the $T_{\mathrm c}(H)$ curves for Nb/Co(30 nm)/Py(45 nm)/Nb thin films along with the magnetic hysteresis curve. We observe a square \textit{M}(\textit{H}) loop for Co(30 nm)/Py(45 nm) bilayer. The square nature of \textit{M}(\textit{H}) of Co(30 nm)/Py(45 nm) bilayer suggests that there is a single magnetization switching and hence, no non-collinear magnetic structure forms in this bilayer. We notice that the transition temperature remains almost constant for all field values lying in the range from positive saturation to negative saturation. The domains switch from positive saturation to negative saturation at a single magnetic field as shown in Fig. 4 and hence, no domain wall formation takes place. Therefore, no suppression in $T_{\mathrm c}$ was observed due to the intrinsic domain wall stray field. Furthermore, no enhancement in $T_{\mathrm c}$ was observed unlike the previous reports\cite{21,22}. This may be due to the absence of non-collinear magnetic structure. Comparing this result with that shown in Fig. 2(b), we propose that the non-collinear structure formed in the Co/Py bilayer may be the source of observed $\bigtriangleup T_{\mathrm c} $ in Fig. 2(a). Moreover, non-collinear magnetic structure has been demonstrated as the key ingredient for the generation of triplet Cooper-pairs from singlet Cooper-pairs\cite{3}. Therefore, we propose that this observation of a recovery of $T_{\mathrm c}$ ($\bigtriangleup T_{\mathrm c} $) in Fig. 2(a) is due to odd-triplet superconductivity in these structures which can be easily modulated with a small applied magnetic field.     Since there is no transport current in our $T_{\mathrm c}$ measurements, there is no lorentz force on the external field induced vortices. Therefore, the effect of vortices is unlikely to play a role in determining the $T_{\mathrm c}$.  Magnetic vortices due to applied fields should affect the spring and non-spring samples equally. Therefore, the comparison of $T_{\mathrm c}$(H) curves in the two cases would not be affected.
  		\newline

	In summary, we have investigated the superconductor proximity effect in Nb(55 nm)/Co(30 nm)/Py(\textit{x})/Nb(55 nm) multilayers with a Py thickness (x) varying from 30 nm to 90 nm. $T_{\mathrm c}$(\textit{H}) curves are investigated from \textit{M}(\textit{H}) loops. It was found that the $T_{\mathrm c}$ of Nb decreases sharply due to the stray field effect of intrinsic domain walls of Py. In the non-collinear range of Co and Py moments, $T_{\mathrm c}$ recovers gradually until both Co and Py layers switch to a parallel magnetization state. As $T_{\mathrm c}$ has been extracted from \textit{M}(\textit{T}) curves, we interpret this recovery of $T_{\mathrm c}$ in the inhomogeneous range  due to singlet-triplet pair conversion at the S/F interface as a result of non-collinear magnetic structure. This recovery in $T_{\mathrm c}$ of about 400 mK cannot be ascribed to a magnetic layer stray field cancellation due to a relative angle between Co and Py moments. In that case, the $T_{\mathrm c}$ in the reversible range of magnetic field should be higher than the $T_{\mathrm c}$ of the saturation range of magnetic field, where the layer stray field is maximum. In this work we have also demonstrated the tuning of $T_{\mathrm c}$ in the magnetically reversible non-collinear range of Co/Py bilayer. Moreover, we have studied the $T_{\mathrm c}$(\textit{H}) curves in Nb/Co(30 nm)/Py(45 nm)/Nb thin films where Co/Py bilayer acts as a single magnetic film and no non-collinear structure forms. No change in $T_{\mathrm c}$ was observed in the field range from positive saturation to negative saturation. This shows that $T_{\mathrm c}$ is dependent on magnetic structure. For a collinear magnetic structure, no changes in $T_{\mathrm c}$ are observed. We show that signatures of singlet-triplet pair conversion are visible in magnetization measurements and Nb/Co/Py/Nb multilayer systems are robust, tunable systems for manipulating singlet to triplet pair converion. This may open up the possibility of an external field-tunable triplet switching mechanism relevant to superconducting spintronics in Josephons junctions. 
	
	We acknowledge funding from National Institute of Science Education and Research(NISER), HBNI, Department of atomic energy(DAE), DST-Nanomission (SR/NM/NS-1183/2013) and DST SERB (EMR/2016/005518), India. J.W.A. Robinson acknowledges funding from the Royal Society and the EPSRC through a Programme Grant (EP/M50807/1) and International Network (EP/P026311/1).

\end{document}